\documentclass[12pt]{article}
\usepackage{amsfonts}
\usepackage{bbm}
\usepackage{graphicx}
\usepackage{sectsty}
\sectionfont{\large}
\usepackage{appendix}
\usepackage{float}
\usepackage{url}
\begin{document}
\title{On the Foundations of the Theory of Evolution}
\author{\normalsize Diederik Aerts$^1$, Stan Bundervoet$^2$, Marek Czachor$^3$, Bart D'Hooghe$^2$, \\
\normalsize Liane Gabora$^4$, Philip Polk$^5$ and Sandro Sozzo$^2$ \\ \\ 
        \small\itshape \vspace{-0.1 cm}
        $^1$ Center Leo Apostel for Interdisciplinary Studies \\
        \small\itshape \vspace{-0.1 cm}
        Department of Mathematics and Department of Psychology \\
       \vspace{-0.1 cm} \small\itshape
         Brussels Free University, Brussels, Belgium \\
       \vspace{-0.1 cm} \small
        Email: \url{diraerts@vub.ac.be} \vspace{0.2 cm} \\ 
        \small\itshape \vspace{-0.1 cm}
        $^2$ Center Leo Apostel for Interdisciplinary Studies \\
       \vspace{-0.1 cm} \small\itshape
         Brussels Free University, Brussels, Belgium \\
       \vspace{-0.1 cm} \small
        Emails: \url{stan.bundervoet@gmail.com,bdhooghe@vub.ac.be,ssozzo@vub.ac.be} \vspace{0.2 cm} \\         \small\itshape \vspace{-0.1 cm}
        $^3$ Department of Theoretical Physics and Quantum Information \\
       \vspace{-0.1 cm} \small\itshape
         Politechnika Gdanska, Poland \\
       \vspace{-0.1 cm} \small
        Email: \url{mczachor@pg.gda.pl} \vspace{0.2 cm} \\
        \small\itshape \vspace{-0.1 cm}
       $^4$ Department of Psychology and Computer Science \\
       \vspace{-0.1 cm} \small\itshape
         University of British Columbia, Okanagan, Canada \\
       \vspace{-0.1 cm} \small
        Email: \url{liane.gabora@ubc.ca} \vspace{0.2 cm} \\
        \small\itshape \vspace{-0.1 cm}
        $^5$ Emeritus, Department of Biology \\
       \vspace{-0.1 cm} \small\itshape
        Brussels Free University, Belgium \\
       \vspace{-0.1 cm} \small
        Email: \url{philip.polk@pandora.be} \vspace{0.2 cm} \\
         }
\date{}
\maketitle
\begin{abstract}
\noindent 
Darwinism conceives evolution as a consequence of random variation and natural selection, hence it is based on a materialistic, \emph{i.e.}~matter-based, view of science inspired by classical physics. But matter in itself is considered a very complex notion in modern physics. More specifically, at a microscopic level, matter and energy are no longer retained within their simple form, and quantum mechanical models are proposed wherein potential form is considered in addition to actual form. In this paper we propose an alternative to standard Neodarwinian evolution theory. We suggest that the starting point of evolution theory cannot be limited to actual variation whereupon is selected, but to variation in the potential of entities according to the context. We therefore develop a formalism, referred to as \emph{Context driven Actualization of Potential} (\emph{CAP}), which handles potentiality and describes the evolution of entities as an actualization of potential through a reiterated interaction with the context. As in quantum mechanics, lack of knowledge of the entity, its context, or the interaction between context and entity leads to different forms of indeterminism in relation to the state of the entity. This indeterminism generates a \emph{non-Kolmogorovian} distribution of probabilities that is different from the classical distribution of chance described by Darwinian evolution theory, which stems from a `actuality focused', \emph{i.e.} materialistic, view of nature. We also present a quantum evolution game that highlights the main differences arising from our new perspective and shows that it is more fundamental to consider evolution in general, and biological evolution in specific, as a process of actualization of potential induced by context, for which its material reduction is only a special case.
\end{abstract}

\section{Introduction\label{intro}}
According to \emph{Neodarwinian Synthesis}, evolution is a consequence of random variation and natural selection of the `fittest' \cite{d59,g02}. In living matter, mutations or any evolutionary novelties randomly arise --- `variation' --- and consequently beneficial mutations are preserved because they aid survival, while others will be destroyed because they withhold survival --- `natural selection'. Some authors have opposed this view on the basis that variation is not merely random but in a large part determined by natural law, \emph{e.g.} Denton, Marshall and Legge argument extensively on the limited variability of protein folds \cite{dml02}. Lately, with the upcoming of new experimental techniques and disciplines as \emph{Evo--Devo}, more and more similar critiques are given paying attention mainly to the `lawlike' developmental constraints on variation and evolution \cite{a04}. It is suggested that the role of natural selection in evolution is significantly smaller than was previously accepted by biologists after Darwin. Natural law in this view has a much greater impact on evolution; it can sometimes be seen as the primary factor in explaining evolutionary change. Notwithstanding the fact that this view has its merits in pointing out the narrowness of the Darwinian account vis a vis physical science, it is, like its opponent, essentially based on a classical, determinist conception of the physical world. This paper proposes a different critique of Neodarwinian theory, which derives instead from a nonclassical and nondeterminist view of physics, as developed recently in quantum mechanics. If the notion of variation is examined carefully, one realizes that what is naturally selected for in the Neodarwinian view are essentially forms of concrete and actual matter. We present a more general view in which forms of potentiality coexist with forms of actuality. We will see that the presence of potentiality states points to a non-Kolmogorovian probability structure at the basis of the context--entity interaction in evolution, which makes possible different pathways of evolution than were allowed for before. 

Since the birth of quantum mechanics as a physical theory describing the behaviour of microscopic particles, the notion of `potentiality state' with respect to different noncompatible contexts has gained importance\cite{adh10}. When the mathematical foundations of quantum mechanics were examined more closely, it turned out that the quantum mechanical nature of the entities is not just due to their microscopic size, but rather can be explained by the presence of potentiality states (superposition states in standard quantum mechanics) and the way in which the context of a measurement, when applied to a quantum entity in a potentiality state, actualizes one of the possible outcomes with a certain probability and induces a state transition toward an eigenstate of the observed outcome. This indeterministic and noncontinuous state transition due to interaction with the measurement context is essentially different from the `free evolution' when no measurement context is applied. In fact, it is often stated that these two types of evolution are contradictious with each other, resulting in many studies of so-called `quantum paradoxes' suggesting that quantum theory is `incomplete' and should be replaced by a more general theory in order to describe both kinds of evolution in a unified way (for a more detailed discussion, see \cite{o94}.

The CAP formalism resolves this basic problem in a specific way. CAP describes the evolution of an entity in a general way as a process of \emph{actualization of potential under influence of context}. The context, as we conceive it, is not limited to what in quantum mechanics is called a `measurement'. If the entity under consideration is in `free evolution', and hence `under influence of the rest of the universe', we consider, within the CAP approach, this `rest of the universe' as the context. This means that if the entity is a microscopic quantum entity, its evolution under the context `rest of the universe' is described by Schr\"{o}dinger's equation. The evolution is in this case continuous and deterministic. However, if a measurement is performed on a quantum entity, then the entity evolves according to the projection postulate, by which the measurement interaction induces a state transition of the entity toward an eigenstate of the observed outcome. In this way, the CAP formalism recovers both types of quantum evolution, the Schr\"{o}dinger one, and the projection one, as special cases of a global evolution.

A \emph{potentiality state} is defined with respect to a given context, \emph{e.g.} the context in which a position measurement is performed, or the context in which a momentum measurement is performed. Because of lack of knowledge about the precise nature of this contextual interaction, one cannot always infer from knowledge about the state of the entity which outcome will occur for a measurement. The resulting quantum probabilities have a very different structure than those encountered in classical experiments. Indeed, in classical theories probabilities of measurement outcomes can always be explained as due to a lack of knowledge about the state of the entity. This leads to a probability model over the set of outcomes, which satisfies the axioms of Kolmogorov \cite{k50}. Quantum probabilities are not due to a lack of knowledge about the state of the entity, but reflect the presence of potentiality states and an uncontrollable interaction between the measurement context and the entity which forces the entity to make a state transition toward an eigenstate of the actualized outcome (an eigenstate for an outcome of an experiment is a state for which this outcome occurs with certainty, whenever one chooses to perform this experiment). As a consequence, the resulting probability model over the set of outcomes has a nonclassical or non-Kolmogorovian structure, the reason being that some of Kolmogorov's axioms are violated.

The process of actualization of potential by means of context is not limited to evolutions within the micro-world described by standard quantum mechanics. It also appears readily in other fields of science, for example in biology \cite{ga05a,ga07}, cognition \cite{badh06}, in social sciences \cite{aa94}, in language \cite{acdh05}. For the sake of completeness, let us sketch the content of this paper. We summarize in Sec. \ref{cap} the main features of the CAP formalism pointing out, in particular, that it provides a more general scheme for evolution theory embodying the Neodarwinian paradigm. Then, we present in Sec. \ref{quant_evo} a simple example that shows the physical differences between the evolution of a classical entity and that of a quantum entity. Finally, we introduce in Sec. \ref{quant_game} a quantum-like evolution game in which evolution along potentiality states produces new states unrealizable by classical means. Thus fact provides a strong support to regard Darwinian evolution, based on natural selection of materially actualized states, as a special case of a more general type of evolution, in which potentiality states and contexts different from the material context play a fundamental role. 

\section{The CAP Formalism\label{cap}}
We resume in this section the essentials of the \emph{CAP formalism} that will be needed to attain our results in the next sections. We refer to \cite{ga05a,ga07} for a detailed discussion of our approach.

The CAP formalism provides a mathematical framework that describes how entities evolve, that is, change over time under the influence of a context. This theoretical framework is based on a generalization of quantum mechanics that was elaborated by ourselves with the aim of describing any kind of contextual interaction between entities (\emph{SCOP formalism}). In the SCOP formalism each (physical, biological, cognitive, etc.) entity is associated with a set $\Sigma$ of states, a set $\mathcal M$ of contexts, a set $\mathcal L$ of properties, and two functions $\mu$ and $\nu$. The function $\mu$ is defined as $\mu: \Sigma \times {\mathcal M} \times \Sigma \longrightarrow [0,1]$, $(q,e,p) \mapsto \mu(q,e,p)$, where $\mu(q,e,p)$ is the probability that the entity in state $p$ changes to state $q$ under the influence of the context $e$. Hence $\mu$ describes the structure of the contextual interaction of the entity under study. The function $\nu$ is defined as $\nu: \Sigma \times {\mathcal L}  \longrightarrow [0,1]$, $(p,a) \mapsto \nu(p,a)$, where $\nu(p,a)$ is the probability that the entity has property $a$ when it is in state $p$. Hence $\nu$ describes the internal structure of the entity or, better, how properties depend on the different states the entity can be in.

There are situations in which we do not have a perfect knowledge of the state of the entity, the context, or the interaction between them, hence the CAP formalism has to cope with nondeterminism. Let us consider an entity in a state $p(t_i)$ under the influence of a context $e(t_i)$ at time $t_i$. If we know with certainty that $p(t_i)$ changes to state $p(t_f)$ at time $t_f$, we refer to the change of state as \emph{deterministic}. If, on the contrary, an entity in a state $p(t_i)$ under the influence of a context $e(t_i)$ at time $t_i$ may change to any state in the set $\left\{ p_1(t_f), p_2(t_f), \ldots , p_n(t_f), \ldots \right\}$, we refer to the process as \emph{nondeterministic}. Nondeterministic change can then be divided into two kinds. In the first, the nondeterminism derives from a lack of knowledge about the state $p(t_i)$ of the entity. In this case the change of state is intrinsically deterministic but, since we lack knowledge about what happens at the deeper level, a model describing what we know is nondeterministic. This kind of nondeterminism is modeled by a stochastic theory that makes use of a probability structure that satisfies Kolmogorov's axioms. Another possibility is that the nondeterminism arises through lack of knowledge about the context $e(t_i)$, or how that context interacts with the entity. It has been proven that in these cases the stochastic model to describe this situation necessitates a non-Kolmogorovian probability model because a Kolmogorovian probability model cannot be used. Since the entity has the potential to change to many different states (given the various possible states the context could be in, since we lack precise knowledge of it), we say that it is in a \emph{potentiality state} with respect to context. We stress that a state is only a potentiality state in relation to a certain (incompletely specified) context. Moreover, we say that the entity is in an \emph{eigenstate} $p(t_i)$ with respect to context $e(t_i)$ if $p(t_i)$ does not change under $e(t_i)$ ($\mu(p(t_i),e(t_i),p(t_i))=1$). This discussion shows that the CAP formalism incorporates several processes of change that may differ with respect to the degree to which they are sensitive to, internalize, and depend upon a particular context, and whether the change of state is deterministic or nondeterministic.

Let us now study the dynamics of entities modeled by SCOP, that is, how CAP concretely works. A dynamical model for CAP can be represented as follows: at time $t_i$ we have a SCOP $(\Sigma, {\mathcal M}, {\mathcal L}, \mu, \nu)(t_i)=(\Sigma(t_i), {\mathcal M}(t_i), {\mathcal L}, \mu, \nu)$ where $\Sigma(t_i)$ is the set of states of the entity at time $t_i$ and ${\mathcal M}(t_i)$ is the set of relevant contexts at time $t_i$. $\mathcal L$ is independent of time, since it is the collection of properties of the entity under consideration. Properties themselves do not change over time, but their status of \emph{actual} ($\nu(p(t_i),a)=1$) or \emph{potential} ($\nu(p(t_i),a) \ne 1$) can change with the change of the state of the entity. Let us now consider four times $t_0, t_1, t_2, t_3$. It may happen that a context $e(t_0)$ has an influence on the entity in state $p(t_0)$ in such a way that the entity can change into one of the states of the set $ \left\{ p_1(t_1), p_2(t_1), p_3(t_1), \ldots, p_n(t_1)\right\}$. This is an example of a general nondeterministic type of change. Similar types of changes can be due to the influence of contexts $e(t_1)$ (the entity can change into one of the states of the set $ \left\{ p_1(t_2), p_2(t_2), p_3(t_2), \ldots, p_m(t_2)\right\}$) and $e(t_2)$ (the entity can change into one of the states of the set $ \left\{ p_1(t_3), p_2(t_3), p_3(t_3), \ldots, p_k(t_3)\right\}$). What is important to observe is that the actual change taking place is a given path, \emph{e.g.} $p_1(t_1), p_4(t_2), p_3(t_3)$, but the states not touched by this path remain of influence for the overall pattern of change, since they are potentiality states. The states $p_0(t_0)$, $p_1(t_1)$, $p_4(t_2)$, $p_3(t_3)$, \ldots constitute the trajectory of the entity through state space, and describe its evolution in time. Thus, the general evolution process is broadly construed as the incremental change that results from recursive, context-driven actualization of potential. A model of an evolutionary process may consist of both deterministic segments, where the entity changes state in a way that follows predictably given its previous state and/or the context to which it is exposed, and/or nondeterministic segments, where this is not the case. With a generalized quantum formalism such as SCOP it is possible to describe situations with any degree of contextuality. In fact, classical and quantum come out as special cases: quantum at one extreme of complete contextuality, and classical at the other extreme, complete lack of contextuality. This is why it lends itself to the description of context-driven evolution.

The CAP framework unifies physical, biological and cultural evolution and allows one to solve some specific problems of these disciplines. In particular, when applied to biological evolution the CAP formalism has been fruitful for illustrating in broad terms how unusual Darwinian evolution is, and clarifying in what sense change of state of living organisms is (and is not) Darwinian. As we have observed above, when change of state of an entity depends on how that entity interacts with a context, the resulting probabilities can be non-Kolmogorovian, and the appropriate formalisms for describing this process are either the quantum formalisms or mathematical generalizations of them. This means that our critique within CAP on the Neodarwinian Synthesis, where the process of evolution is narrowed down to the interplay of variation and selection, is structural. A process of interplay of variation and selection as imagined within the Neodarwinian Synthesis will lead to an underlying probability model which is Kolmogorovian, and in this sense cannot take into account the influence of context as conceived within CAP. In practice it means that the narrowed down evolution process, as within the Neodarwinian Synthesis, where only variation and selection play, neglects what happens in the realm of the potential states of the considered entities. It only considers actualized entities, and their actualized interactions, and believes that all of evolution will be steered by these actualized entities, and their actualized interactions. As we have seen above, within CAP also potential states of entities, and potential interactions between these entities will play a role in the process of evolution. In this respect CAP is a generalization of the Neodarwinian process, and in this respect furthers the work of scientists who saw the importance of non-Darwinian processes in biology. We do not insist on this point, for the sake of brevity (the interested reder can see, \emph{e.g.} \cite{ga05a,ga07,acgp06}). We instead show in the next sections by means of concrete examples how relevant potentiality states may be for physical (and non-physical) evolution.  

\section{Classical versus Quantum Evolution\label{quant_evo}}
Let us illustrate the main differences between classical and quantum evolution and the relevance of potentiality states with a simple example. Consider a point particle, \emph{e.g.} a bullet, which is fired with a certain momentum straight toward a plate containing a slit. This defines a measurement of position ($x$ and $y$ coordinates of the slit on the plate). If the particle hits the plate, it gets destroyed and the measurement yields a negative result. In thise case we say that `the bullet didn't pass the plate'. If however, the particle arrives at the plate at the spot $(x_0,y_0)$ where the slit is located, it will pass the plate without any problem, yielding a positive result for the `plate-test'. In this case we say that `the bullet did pass the plate via the slit at point $(x_0,y_0)$'. Next, a second position measurement is performed by putting another plate parallel with the first one, such that the slit is located a bit to the left of the first slit. Since the bullet travels straight toward the two plates and the positions of the two slits are different, classical physics predicts that no classical bullet can pass this configuration of two plates with one slit or, equivalently, the probability that a classical bullet passes the configuration is zero.

In quantum mechanics, the situation is different. Position and momentum are complementary observables, \emph{i.e.} if the entity is in an eigenstate of one observable, it necessarily will be in a superposition state with respect to the second observable. This is due to the fact that the position and momentum operators do not commute, implying they do not have a common set of eigenvectors, which means that it is impossible to be in an eigenstate for both reference frames at the same time. Since one can only claim that a certain observable has a value if and only if the entity is in an eigenstate of that observable, it follows that for a quantum entity it is impossible that at the same time both position and momentum have a value. In mathematical terms this is reflected as follows. The states of the quantum entity are represented in an abstract vector space, a complex Hilbert space. Depending on which context one chooses to actualize the potential (position or momentum) one has to choose either a set of eigenstates of position to express the state of the entity, or alternatively a set of eigenstates of momentum. It is a fact of quantum theory that no common set of eigenstates exists for these two observables. Hence, if the entity is in an eigenstate of position, then it is necessarily in a potentiality state with respect to a context defined by a momentum measurement, and vice versa. By referring to the bullet example considered above, this result entails that after the first position experiment the quantum entity has changed to an eigenstate of position, corresponding with the position of the slit in the first plate. Hence immediately after this first measurement, the state of the entity will be an eigenstate of position. However, as soon as the particle has passed the first slit, its wave function will spontaneously spread out over space, resulting in a superposition state of position eigenstates. In a sense, one could interpret this as the natural evolution of the state of the entity from a limited class of eigenstates of position toward the larger general class of `spread out states'. It follows that the probability that the quantum bullet will pass the second slit is non-zero, at variance with the classical bullet. This example illustrates how by passing from one context and set of eigenstates (position) via a potentiality state (the spread out wavefunction between the two plates) toward the final state which is again an eigenstate of position (after the second plate), the entity can evolve from an eigenstate of position corresponding with the first slit toward a different eigenstate of position defined by the second slit, which is not possible in the classical case.

We have shown in this section that new possibilities arise for the evolution of an entity which are classically forbidden, if evolution itself is regarded as an actualization of potential induced by the interaction between the entity and its relevant context. This achievement will be strengthened in Sec. \ref{quant_game} by elaborating even more convincing examples.

\section{Children Playing a Quantum-like Evolution Game\label{quant_game}}
In this section we propose an example of a entity which evolves from one eigenstate to another eigenstate with respect to a certain context through a potentiality state, which transcends the mere `random variation and selection upon fitness' scheme. We consider 6 objects: two pair of scissors, one blue (Sb) and one red (Sr), two rocks, one blue (Rb) and one red (Rr), and two pieces of paper, again one blue (Pb) and one red (Pr). The entity consists of two boxes, each containing exactly two objects. This means that there are two objects left, which can be put in `a bag of spare parts'. In order to play the quantum-evolution game, we need the assistance of three children, let's call them Niels, Katia and Jonas. Niels is playing with the objects in the left box, while Katia is playing with the objects in the right box. The bag of spare parts is for Jonas, who can pick a piece out of this bag to play the game together with Niels or Katia. Two classes of measurements can then be performed on the entity.

A measurement `matching blue objects' is performed as follows. The observer (Niels or Katia) wants to have two blue objects in his/her box. First, the observer looks at the pieces in his/her box. If they both are blue (\emph{e.g.}, Rb, Sb), then the observer is satisfied with this configuration and the measurement yields `yes', and the state of the entity is unchanged. If they both are not blue (hence both red) then the measurement yields `no' and the red objects are left in the box as they are. If however one of the two pieces is blue, the measurement proceeds as follows. The observer asks Jonas to pick randomly one of the two objects from his `bag of spare parts'. If it is blue, (s)he replaces the red object from his/her box with the blue object from Jonas' bag. If it is not blue (hence red), (s)he puts this red piece back in Jonas' bag and after some time decides to give it a second try, and Jonas takes again at random an object from his bag. If it is blue, then it is used to replace the red object, but if it is again red, then the observer gives up trying to `match blue objects' and accepts that there are two objects of the other colour, \emph{i.e.} the blue object is replaced with the red one. As such, (s)he doesn't have two matching blue objects, but at least there are two matching colour objects (red objects). Obviously, since there are not two matching
blue objects, the `matching blue objects' measurement yields outcome `no'.

A second class of measurements is based on the children's game `paper, scissors, rock', in which two players pick at random one of the three possibilities, and then check their moves. If they both made the same choice, it is a draw. If however there is a mismatch, then the rules are as follows: paper defeats rock, rock defeats scissors, but scissors defeat paper. The measurement `matching pairs of objects' is defined as follows. If the two objects in the box are the same (\emph{i.e.} have the same shape), then the measurement yields outcome `yes', and the state is left as it is (hence the box is in an eigenstate of `both objects having the same shape'). If not, the observer asks Jonas to select at random one of the two objects from his `bag of spare parts' and one of the two objects in the observer's box. Next, the two selected objects play the `paper, scissors, rock' game. If the selected object from Jonas' bag wins, it replaces the selected object from the box. Otherwise (in case of a draw or loss), the selected object from Jonas' bag plays the `paper, scissors, rock' game against the second object from the box. Again, if the object from Jonas' bag wins, it replaces the object in the box. If the object selected from Jonas' bag loses or makes a draw against both objects from the box, the measurement continues as Jonas takes the second object from the bag and plays the `paper, scissors, rock' game, applying the same rules as he did for the first object from the bag. So if both objects from Jonas' bag lose or draw against both objects of the box, the observer will keep the same (different) objects in his/her box and the measurement yields `no', \emph{i.e.} `(s)he did not succeed in matching two objects of the same shape in the box'. In the other case, (s)he has replaced an object from his/her box with an object from the bag; if the two objects in his/her box now have the same shape, the measurement yields `yes' (`I have now two objects with the same shape in my box'); if not, the measurement yields `no'.

For clarity, let us briefly discuss the possible results of this measurement procedure, abbreviating  P(aper), R(ock) and (pair of) S(cissors) with their initials:
\begin{itemize}
\item Initial state of the box: PP, RR or SS: the measurement always yields outcome `yes'.
\item Initial state of the box: PR, PS or RS. Let us consider the case PR (such that the different possible configurations of objects in the bag is given by PR, PS, RS or SS). The discussion for the two other cases (\emph{i.e.} the box containing PS or RS) is analogous (under cyclic permutation of P, R and S).
\begin{enumerate}
\item[(A)] Box: PR (bag: PR): R from the bag always loses or draws against P or R from the box, P wins against R from the box, so the final configuration of the box is PP (bag: RR); hence the outcome is `yes'.
\item[(B)] Box: PR (bag: PS). If P is selected from the bag, it draws against P from the box, but wins against R, so the new state becomes PP (bag RS) ; hence `yes'. If however S is selected, it loses from rock but wins against P, so final state of the box is SR (bag: PP), and outcome `no'.
\item[(C)] Box: PR (bag: RS). R from the bag always loses or makes draw, S loses against R but wins against P, so the new state of the box is SR (bag: RP), outcome `no' is given to the experiment.
\item[(D)] Box: PR (bag: SS). S wins against P, new state SR (bag : PS) but since these are different objects, the measurement yields outcome `no'.
\end{enumerate}
\end{itemize}

One can show that for this entity and set of experiments one can derive a violation of Bell's inequalities \cite{b64}, which proves that this entity cannot be described within a Kolmogorovian \emph{i.e.}, classical scheme.

Next, let us look at the possible evolutions for a compound entity of two boxes with actualization driven under the context of `matching colour' measurements. Let the entity be prepared in the configuration (Pb, Sr) for Niels' box, and (Pr, Sb) for Katia's box. Hence, Jonas' `bag of spare parts' contains (Rb, Rr). Now the `matching blue objects' context can be applied, such that after the measurement two possibilities can be realized: either Niels succeeds in getting two `matching blue' objects such that consequently Katia has two red objects, or else Katia succeeds in obtaining two blue objects, and Niels ends up with two red objects. When they continue playing the `colour matching game', they find that they cannot change their state with respect to colour: {\it e.g.} in the first case, Niels stays in an eigenstate of having two blue objects, and Katia stays in an eigenstate of `red objects'. So in this scheme, no further `evolution' is possible.

Next, let us consider the ``matching shapes through the `paper, scissors, rock' game'' measurements. Let Niels be in an eigenstate of blue objects, \emph{e.g.} (Pb, Rb) and Katia in an eigenstate of red objects (Sr, Pr). Consequently, Jonas' bag contains (Rr, Sb). If Niels performs the `matching shapes measurement' such that Rr is chosen from the bag, it will always lose (Pb) or make a draw (Rb) and be put back into the bag. However, if Sb is selected, it can win from Pb, and replace this object in Niels' box. Hence the new state of Niels' box becomes (Sb, Rb) such that the bag contains (Pb, Rr). If we now again apply the `matching shapes' context, then if Rr is selected from the bag, this object defeats Sb, such that Niels' box changes into the new configuration (Rr, Rb), which is an eigenstate of `matching shapes experiment', but a potentiality state with respect to the `matching colour' context. Note that the bag now contains (Pb, Sb). Next, Katia could apply the `matching shapes' context on her box. Pb always loses (Sr) or makes a draw (Pr) if played against an object from Katia's box, but if Sb is selected from the bag, it defeats Pr, such that the new configuration of Katia's box becomes (Sr, Sb), which is a potentiality state with respect to the `matching colours' context. In this way, the compound entity of Niels' and Katia's box is brought into a potentiality state with respect to the `matching colours' experiment. Next, let us see what happens if a `matching colours' context is applied. Let Katia perform the measurement `matching colour blue'. Her box contains (Sr, Sb), Jonas' bag of spare parts contains (Pr, Pb) and Niels' box contains (Rr, Rb). Hence with non-zero probability Katia could change her configuration into (Sb, Pb), which is an eigenstate of colour blue. Following, Niels can apply the `matching colours context' to obtain his new stable configuration, namely of red objects (Rr, Pr) or (Rr, Sr), depending on which object was selected from the bag.

The above toy model demonstrates that, by applying a different context and considering potentiality states of the entity with respect to the initial context, new ways of evolution are possible. In particular, the box of Katia changed from a `red' eigenstate with respect to the context of `matching colours' toward a `blue' eigenstate by considering intermediate potentiality states, realized by actualizing under the ``matching shapes through the `paper, scissors, rock' game'' context. Evolution is thus realized by actualizing the potential under different contexts. CAP has major implications for evolution theory, which up till now only considered eigenstates of the `selection upon fitness' context, while potentiality states were excluded. If we take notice of the fact that the physics underlying Neodarwinian evolution theory has entirely changed, it is no surprise that evolution theory will have to change. A first step in that direction is to provide a broader view on evolution that contains Darwinian evolution (defined by variation and selection upon fitness with respect to a certain context) as a special case, in the sense that all intermediate states in the evolution of the entity are classical states, and eigenstates of the considered `selection upon fitness' context, while potentiality states are excluded, but that contains also evolutionary processes going through potentiality states toward different contexts and bringing forth entirely different situations.

\section{Conclusions\label{conc}}
As we have argued using the model of Niels, Katia and Jonas playing the quantum evolution game, if evolution is realized by inducing different, noncompatible contexts to the entity, then evolution along potentiality states could result in new states unrealizable by classical means. Hence pathways for evolution exist which do not exist in the classical case. In this approach, Darwinian evolution, based on natural selection of materially actualized states, can be regarded as a special case of a more general type of evolution, in which potentiality states and contexts different from the material context play a fundamental role. The framework of CAP also makes it possible to view biological evolution as one of the possible types of evolution, and for example cultural evolution as another type of evolution within the framework \cite{ga05a,ga07}. Also, CAP constitutes the first step toward the elaboration of an evolution theory which can cope with more recent developments in physics and with the idea of a physical aspect of evolution compatible with these modern physical theories (\emph{i.e.} quantum mechanics). A promising future discussion could be made on the relation between quantum indeterminism of potentiality states and the presumed randomness of variation, which lies at the basis of the Darwinian theory. As philosopher of science Karl Popper pointed out, randomness, or indeterminism of results, is a necessary condition for natural selection, which he regards as a form of downward causation, to have any explanatory power \cite{p82}. But until recently, no `hard' sciences had come up with a founded view on indeterminism. Quantum mechanics is probably the first candidate to serve this role. We conclude that it is more fundamental to consider evolution in general, and hence also biological evolution in specific, as a process of `context driven actualization of potential', for which its material reduction is a reduction within one specific context, namely the context of these moments of evolution which are actual within this materialistic context.

\end{document}